\documentclass[aps,prb,preprint,groupedaddress,showpacs]{revtex4}
\usepackage{amsmath,graphicx,tabularx,subfigure,bm}

\def\bra#1{\mathinner{\langle{#1}|}}
\def\ket#1{\mathinner{|{#1}\rangle}}

\def\e{\epsilon}
\def\t{\theta}
\def\g{\gamma}
\def\o{s}
\def\s{\sigma}
\def\st#1{\mbox{\scriptsize{#1}}}

\def\kf{k_{\st{F}}}
\def\vf{v_{\st{F}}}

\begin{document}
\title{Metal-Semiconductor Transition and Fermi Velocity Renormalization in Metallic Carbon Nanotubes}
\author{Yan Li}
\email{yanli@uiuc.edu}
\author{Umberto Ravaioli}
\affiliation{Beckman Institute for Advanced Science and Technology, University of Illinois at Urbana-Champaign, Urbana, Illinois 61801}
\author{Slava V. Rotkin}
\affiliation{Department of Physics, Lehigh University, Bethlehem, Pennsylvania 18015}
\date{\today}

\begin{abstract}
Angular perturbations modify the band structure of armchair (and other
metallic) carbon nanotubes by breaking the tube symmetry and may
induce a metal-semiconductor transition when certain selection rules
are satisfied. The symmetry requirements apply for both the nanotube
and the perturbation potential, as studied within a nonorthogonal
$\pi$-orbital tight-binding method. Perturbations of two categories
are considered: an on-site electrostatic potential and a lattice
deformation which changes the off-site hopping integrals.  Armchair
nanotubes are proved to be robust against the metal-semiconductor
transition in second-order perturbation theory due to their high
symmetry, but can develop a nonzero gap by extending the perturbation
series to higher orders or by combining potentials of different
types. An assumption of orthogonality between $\pi$ orbitals is shown
to lead to an accidental electron-hole symmetry and extra selection
rules that are weakly broken in the nonorthogonal theory. These
results are further generalized to metallic nanotubes of arbitrary
chirality.
\end{abstract}
\maketitle

\section{Introduction}

The subject of metal-insulator transitions has been studied for
decades.~\cite{MOTT74} It is well understood that a metal-insulator
transition is typically related to the breaking of a specific symmetry
of the system, and carbon nanotubes are particularly interesting to
study, due to their low dimensionality and special helical
symmetry. In this paper we investigate the symmetry breaking in
single-wall nanotubes (SWNTs) due to an external potential which
depends on the angular coordinate along the SWNT circumference. This
perturbation may induce a transition in a SWNT changing the type of
its electronic structure by opening/closing the band gap in the
metallic/semiconducting nanotube.~\cite{DELA98,KANE97, LIU99,PARK99,
LI2003, ROTK2004} It is important for applications in which such gap
engineering can potentially be controlled locally, for instance by the
field of a sharp tip, by a small molecule or by a local gate.

SWNT lattice symmetry depends on two parameters, diameter and
chirality, which determine the type of band structure.~\cite{ROTK2005}
About one-third of possible SWNTs are metallic, with one dimensional
energy subbands crossing at the Fermi level, as confirmed by
experiments.~\cite{DRES2001} In this study, we consider metallic
nanotubes and focus mainly on the special case of armchair SWNT
(A-SWNT), which has a higher symmetry prohibiting an energy gap at the
Fermi level~\cite{DAMN2000} for typical non-chiral perturbations such
as stretching, uniform electric field, impurity potentials and
many-body interactions.  The same perturbations would open a small
``secondary'' band gap in metallic nanotubes of different symmetry.
The high symmetry of A-SWNTs is also responsible for the absence of
back scattering in the conduction channels.~\cite{ANDO} Such ballistic
transport is very attractive for future electronic
applications~\cite{ROTK2005,DRES2001} and a method to control the
conductance of A-SWNTs would be particularly desirable. Different
perturbations have been attempted to modify the electronic structure
of A-SWNTs.~\cite{DELA98,KANE97, PARK99,LIU99, LI2003,LI2004,
ROTK2005,GULS2002a,LU2003,ROTK2004,MEHR2005} Our goal is to
demonstrate, using symmetry arguments, whether a particular
perturbation can open a gap at all and how the gap depends on the
magnitude of the perturbation potential. We will show below that, with
minor exceptions, this cannot be a linear dependence. We call this
transition a ``metal-semiconductor transition'' (MST) since the gap
size is smaller than in typical insulators.~\cite{LI2004}

A nonorthogonal tight binding (TB) approach is used to model the SWNT
electronic structure. Despite its simplicity, the TB approach may
include as much important physics as more sophisticated models with
the right choice of empirical parameters.~\cite {REIC2002} In
addition, it possesses a great advantage for analytical
derivations. Combining the TB approach and the summation of
perturbation series with a group theory technique, it was shown in
previous work~\cite{LI2004} that mirror symmetry breaking is a
necessary condition to mix the two crossing subbands ($\pi$ and
$\pi^*$) and open a band gap in A-SWNTs.  Here, we find several
additional significant results: (1) Due to the high lattice symmetry,
the second order contributions \emph{always} cancel out and no second
order gap opening occurs. One notes that the first order process is
suppressed unless very specific selection rules are satisfied (see
below). (2) Under potentials of a single angular Fourier component,
${\bf V}_q\cos q\t$, the lowest contributing coupling order between
$\pi$ and $\pi^*$ bands, $\mu_0$, is determined by the angular
momentum of the potential, $q$, and the index $n$ of the $(n,n)$
A-SWNT as $\mu_0=2n/\mathrm{gcd} (2n,q) $, in which $\mathrm{gcd}$ is
the greatest common divisor. The band gap opening is proportional to
$|{\bf V}_q|^{\mu_0}$ for small perturbation $|{\bf V}_q|\ll\vf/R$,
where $\vf$ and $R$ are the nanotube Fermi velocity and radius. In a
typical experiment, when the perturbation has a small angular momentum
($q\sim 1$), the coupling order is high ($\mu_0\sim n$) and the gap is
small if any. To observe a linear effect ($\mu_0=1$), a high-$q$
potential must be applied ($q=2n$). The high coupling order can be
reduced by choosing combinations of several angular Fourier components
or different types of perturbation. (3) Additional symmetry of a
particular model may lead to extra selection rules for the band gap
opening. For example, gaping is forbidden for A-SWNTs with even $n$ if
perfect electron-hole symmetry is assumed, as in an orthogonal
basis. When the slight asymmetry between the conduction and valence
bands is included, a gap proportional to the asymmetry parameter
occurs. (4) Significant changes in the A-SWNT density of states (DOS)
is observed even when the gap is absent.  Modification of the low
energy band structure is mostly determined by the second order
perturbation.
The DOS is enhanced near the Fermi level and simultaneously $\vf$
decreases. Peaks of the first pair of van Hove singularities are
brought closer, resulting in a smaller excitation energy between these
subbands.

Our study is restricted to the case of potentials which are uniform
along the tube axis, but certain results are easily generalized for
perturbations with even/odd axial dependence.

This paper is organized as follows. We first formulate the model and
introduce the interaction matrix elements between TB wave functions in
Sec.~\ref{Sec:matrix element}.  Using nearly degenerate perturbation
theory (Appendix~\ref{appendix:NDPT}) and symmetry-based selection
rules, we derive analytically the coupling between $\pi$ and $\pi^*$
subbands of A-SWNTs for both scalar potentials (Sec.~\ref{Sec:scalar
potential}) and tensor potentials (Sec.~\ref{Sec:tensor
potential}). Comparisons are made with numerical results from TB band
structure calculation.  These results are further extended to metallic
nanotubes of arbitrary chirality in Sec.~\ref{Sec:chiral}. Finally, we
summarize our results in Sec.~\ref{Sec:conclusion}.

\section{Model formulation} \label{Sec:matrix element}

\subsection{Perturbation series in the TB description}

The electronic states $\Psi_i$ are obtained within a nonorthogonal
single $\pi$-orbital TB method by solving the stationary Schroedinger
equation
\begin{equation}
  H\Psi_i=E_iS\Psi_i,
\end{equation}
\noindent where $H$ and $S$ are the Hamiltonian matrix and overlap
matrix, respectively. For the sake of simplicity, only nearest
neighbor hopping integral $\g_0=-3.033\hbox{ eV}$ and overlap integral
$\o=0.129$ are considered.~\cite{SAIT98}

The wave function of an unperturbed A-SWNT can be expressed as a
linear combination of the two periodic functions $u_{\xi}({\bf
k})=\frac{1}{\sqrt{N}}\sum_ie^{i{\bf k}\cdot({\bf r}_{i \xi}-{\bf
r})}\varphi({\bf r}-{\bf r}_{i\xi})$, where $\xi=A,B$ label the two
sublattices and $\varphi({\bf r}-{\bf r}_{i\xi})$ is the atomic
orbital function localized at ${\bf r}_{i\xi}$:
\begin{eqnarray}
 \label{wave function}
\Psi_{\s}({\bf k})&=&\frac{e^{i{\bf k}\cdot{\bf
r}}}{\sqrt{2\left[1-s\s |f({\bf k})|\right]}}\left[
e^{i\phi_{\s}({\bf k})}u_A({\bf k})+e^{-i
\phi_{\s}({\bf k})}u_B({\bf k})\right]\nonumber\\
E_{\s}({\bf k})&=&\g_0\frac{-\s|f({\bf k})|}{1-\o\s|f({\bf k})|},
\hspace{,2 in}f({\bf k})=\sum_{\lambda=1}^3e^{i{\bf k}\cdot{\bf
    r}_{\lambda}},
\end{eqnarray}
\noindent where $\s=\pm 1$ denote the conduction and valence bands.
${\bf r}_{\lambda}$'s are the nearest neighbor bond vectors and we
refer to $\lambda=1$ as the circumferential direction of A-SWNTs in
the following.  The wave vector ${\bf k}$ is composed of an axial
component $k_t$ and a quantized circumferential component $k_c=m/R$,
with $m$ the angular momentum. $ 2\phi_{\s}({\bf k})\equiv\arg[-\s
f({\bf k})]$ indicates the phase difference of the coefficients before
$u_A({\bf k})$ and $u_B({\bf k})$, which make a pseudo-spinor (see
also Sec.~\ref{Sec:chiral}). $\phi_{\s}({\bf k})$ is constant through
the whole range of $k_t$ for the two crossing subbands:
$\phi_{\pi}(k_t)=\pi/3$ and
$\phi_{\pi^*}(k_t)=\phi_{\pi}-\pi/2$. Clearly, the $\pi$ and $\pi^*$
subbands are orthogonal due to this phase difference.

Consider a perturbation $H_1$ which is uniform in the axial direction,
then $k_t$ is conserved. To obtain the low energy behavior of $\pi$
and $\pi^*$ subbands, we write the effective $2\times2$ perturbation
Hamiltonian matrix using nearly degenerate perturbation theory (see
Appendix):

\begin{eqnarray}\label{Heff}
  H_{\st{eff}}(k_t)=\left[
\begin{array}{cc}
  E_{\pi}(k_t)+H_{\pi\pi}(k_t)& H_{\pi\pi^*}(k_t) \\
  H_{\pi\pi^*}^*(k_t)& E_{\pi^*}(k_t)+H_{\pi^*\pi^*}(k_t)
\end{array}\right],
\end{eqnarray}
\noindent where the matrix element $H_{\alpha\beta}(k_t)$, with
$\alpha (\beta)=\pi$ or $\pi^*$, can be represented by the sum of the
perturbation series over all possible coupling orders $\mu$ as
\begin{eqnarray}\label{matrix series}
H_{\alpha\beta}(k_t)&=&\sum_{\mu}\sum_{\{\Psi_i\}}H_{\alpha\beta}^{(\mu)}\left(\{\Psi_i\}\right),\hspace{.2
in}\Psi_i\equiv\Psi_{\s_i}(k_t,m_i)\nonumber\\
H^{(\mu)}_{\alpha\beta}(\{\Psi_i\})&=
&\bra{\Psi_{\alpha}}H_1\ket{\Psi_1}\frac{\prod\limits_{i=2}^{\mu-1}\bra{\Psi_{i-1}}
H_1\ket{\Psi_{i}}}{\prod\limits_{i=1}^{\mu-1}(-E_i)}\bra{\Psi_{\mu-1}}H_1\ket{\Psi_{\beta}}.
\end{eqnarray}
\noindent One notes that all intermediate states $\Psi_{i}$ ($i=1\dots
\mu-1$) are different from $\Psi_{\pi}$ and $\Psi_{\pi^*}$ by
definition. Figure~\ref{fig:coupling} illustrates an example of second
order coupling between $\pi$ and $\pi^*$ subbands via four different
paths. Also shown is the phase angle of the intermediate states,
$\phi_{\pm}(k_t, n\pm q)$, relative to $\phi_{\pi}$ and
$\phi_{\pi^*}$.  The relation between these phase angles will be
discussed in detail in the following sections.

\subsection{Interaction matrix element between Bloch states}
Below, we derive the interaction matrix element
$\bra{\Psi}H_1\ket{\Psi'}$ of the angular perturbation of a single
Fourier component, $H_1={\bf V}_q\cos{q(\t-\t_0)}$, where ${\bf V}_q$
could be either scalar or tensor. $\theta_0$
is defined as the minimum angular displacement between the vertical
mirror planes (or glide planes) of the A-SWNT and the mirror planes of
the potential.~\cite{LI2004} Assuming that
$\bra{\varphi_1}H_1\ket{\varphi_2}$ is nonzero only when $\varphi_1$
and $\varphi_2$ are centered on the same atom or nearest neighbor
atoms, the interaction matrix element can be decomposed into an
on-site term and an overlap term:
\begin{eqnarray}\label{Eq.matrix element}
  \bra{\Psi}H_1\ket{\Psi'}&=&\frac{\delta_{m-m',q
 }e^{-iq\t_0}+\delta_{m-m',-q}e^{iq\t_0}}{2\sqrt{(1-s\s|f|)(1-s\s'|f'|)}}\left(\left<H_1\right>_{\st{
 on-site}}+\left<H_1\right>_{\st{overlap}}\right)\nonumber\\
 \left<H_1\right>_{\st{on-site}}&=&\bra{\varphi({\bf r})}{\bf
 V}_qe^{iq\t}\ket{\varphi({\bf r})}\cos(\phi-\phi')\nonumber\\
 \left<H_1\right>_{\st{overlap}}&=&\sum_{\lambda=1}^3\bra{\varphi({\bf
 r}-{\bf r}_{\lambda}/2)}{\bf V}_qe^{iq\t}\ket{\varphi({\bf r}+{\bf
 r}_{\lambda}/2)}\cos\left(\phi+\phi'-\frac{m+m'}{2}\t_{\lambda}-k_tz_{\lambda}\right),
\end{eqnarray}
\noindent where it is used that $\varphi({\bf r})$ is real and
invariant under $\t\leftrightarrow -\t$ inversion. $\delta_{m-m',\pm
q}$ arises from the conservation of the angular momentum. This
conservation law also allows $q$ to differ by multiple of $2n$, an
angular analog of the reciprocal lattice vector, which is not
considered here (see discussion in Sec.~\ref{Sec:chiral}). The
interaction matrix can be further simplified depending on the type of
${\bf V}_q$:
\begin{description}
  \item \emph{Scalar Potential}: ${\bf V}_q=$\emph{const}. Define $
  \bra{\varphi({\bf r})}{\bf V}_qe^{iq\t}\ket{\varphi({\bf r})}=U_q$,
  then
\begin{eqnarray}
 \bra{\varphi({\bf r}-{\bf
r}_{\lambda}/2)}V_qe^{iq\t}\ket{\varphi({\bf r}+{\bf r}_{\lambda}/2)}
&\approx&\frac{1}{2} \o\left[
\bra{\varphi({\bf r}+{\bf r}_{\lambda}/2)}V_qe^{iq\t}\ket{\varphi({\bf r}+{\bf r}_{\lambda}/2)}\right.\nonumber\\
&+&\left.\bra{\varphi({\bf r}-{\bf r}_{\lambda}/2)}V_qe^{iq\t}\ket{\varphi({\bf r}-{\bf r}_{\lambda}/2)}\right]\nonumber\\
&=&\o U_q\cos \frac{q\t_{\lambda}}{2}
\end{eqnarray}
\begin{eqnarray}
 \left<H_1\right>_{\st{overlap}}&=&\frac{1}{4}\o
 U_q\left[e^{i(\phi+\phi')}\sum_{\lambda=1}^3(e^{-im\t_{\lambda}-ik_tz_{\lambda}}+e^{-im'\t_{\lambda}+ik_tz_{\lambda}})+c.c.\right]\nonumber\\
 &=&-\frac{1}{2}\o U_q
 \cos\left(\phi-\phi'\right)\left(\s|f|+\s'|f'|\right),
\end{eqnarray}
\noindent and Eq.~(\ref{Eq.matrix element}) is reduced to
\begin{equation}\label{Eq.scalar}
\langle H_1\rangle_{\st{on-site}}+\langle
H_1\rangle_{\st{overlap}}\approx
U_q\cos(\phi-\phi')\left[1-\frac{1}{2}\o\left(\s|f|+\s'|f'|\right)\right].
\end{equation}

 \item \emph{Tensor Potential}: The Fourier component of a tensor
   potential can be expressed in the second quantization formalism as
   \begin{eqnarray}
\sum_{\lambda=1}^3 \delta  \gamma_{\lambda,q}\sum_{\langle i,j \rangle^{\lambda}}e^{iq(\t_i+\t_j)/2}c^+_ic_j ,
   \end{eqnarray}
\noindent where pairs $\langle i,j\rangle^{\lambda}$ are confined to
first nearest neighbors with bonds along ${\bf r}_{\lambda}$ direction
and $\delta\g_{\lambda,q}$ is the corresponding change of the hopping
integral.  The on-site term is absent while the overlap term is
reduced to:
\begin{eqnarray}\label{Eq.vector}
\left< H_1\right>_{\st{overlap}}&\approx&\sum_{\lambda=1}^3g_{\lambda}(k_t;m,\s;m',\s')\nonumber\\
&=&\sum_{\lambda=1}^3\delta\g_{\lambda,q}\cos\left(\phi+\phi'-\frac{m+m'}{2}\t_{\lambda}-k_tz_{\lambda}\right).
\end{eqnarray}
\end{description}
\noindent Comparing Eq.~(\ref{Eq.scalar}) and Eq.~(\ref{Eq.vector}),
one can conclude that 
the interaction matrix elements from a scalar potential and a tensor
potential have quite different dependence on the phase angle $\phi$
and the quantum numbers $k_t,m$.  We will show in next section that
certain selection rules can be derived for scalar potentials of
general form and also for simple tensor potentials allowing summation
over $\lambda$.

\section{Scalar Potential}\label{Sec:scalar potential}

Assume that a scalar perturbation in the form of $H_1=V_q\cos q(\t-
\t_0)$ is applied to the $(n,n)$ A-SWNT. Using the interaction matrix
element from Sec.~\ref{Sec:matrix element}, we can now derive the
$\mu$-th order perturbation matrix elements within nearly
degenerate perturbation theory:
\begin{eqnarray}
H^{(\mu)}_{\alpha\beta}(\{\Psi_i\})&=&e^{-i\Delta \mu
q\t_0}\left(\frac{U_q}{2}\right)^{\mu}\frac{P_{\alpha\beta}
(\{\Psi_i\})Q(\{\Psi_i\})}{\prod\limits_{i=1}^{\mu-1}(-E^0_i)}\nonumber\\
P_{\alpha\beta}(\{\Psi_i\})&=&\left[\prod\limits_{i=1}^{\mu-1}\cos(\phi_{i-1}-\phi_{i})\right]\cos(\phi_{\mu-1}+\Delta
\mu q\pi/3n-\phi_{\mu})\nonumber\\
Q(\{\Psi_i\})&=&\prod\limits_{i=1}^{\mu}\left[1-\frac{1}{2}\o(\s_{i-1}|f_{i-1}|+\s_i|f_{i}|)\right],
\label{case 1}
\end{eqnarray}
\noindent where subscripts ``$0$'' and ``$\mu$'' correspond to the
initial state $\Psi_{\alpha}$ and final state $\Psi_{\beta}$
respectively, with $m_0=m_{\mu}=n$. We stress that $E^0_i\equiv
E^0_{\s_i}(k_t,m_i)=-\s_i\g_0|f(k_t,m_i)|$, since the factor
$(1-\o\s_i|f_i|)^{-1}$ is canceled by those from the wave
functions. $P_{\alpha\beta}$ is the total phase of the perturbation
term 
and corresponds to inner products of the pseudo-spinors (see
Sec.~\ref{Sec:chiral}). $Q$ corrects for contributions from a nonzero
nearest neighbor overlap. The intermediate states
$\{\Psi_i\equiv\Psi_{\s_i}(k_t,m_i)\}$
satisfy the conservation law of angular momentum and have the
following constraints:
\begin{eqnarray}\label{m_consv}
m_{i-1}-m_{i}&=&\pm q, \qquad i=1,\dots,\mu-1 \nonumber\\
m_{\mu-1}-m_{\mu}&=&\pm q+\hbox{multiple of }2n,
\end{eqnarray}\noindent which leads to
$\sum_{i=1}^{\mu} \left(m_{i-1}-m_i\right)=\Delta\mu q+\mbox{multiple
 of } 2n=0$.  The role of a nonzero $\Delta \mu$ can be seen from the
 extra phase factor $e^{-i\Delta\mu q\t_0}$ in
 $H_{\alpha\beta}^{(\mu)}$ and the corresponding term $\Delta \mu
 q\pi/3n$ in $P_{\alpha\beta}$. Direct evaluation of Eq.~(\ref{case
 1}) with all possible sets of  $\{\Psi_i\}$  is formidable, nevertheless,
 one can get useful information by applying symmetry arguments.

\subsection{Off-diagonal coupling: gapping of A-SWNTs}
We first study the off-diagonal term $H^{(\mu)}_{\pi\pi^*}$ and
replace $m_i$ with $\widetilde{m}_i=2n-m$ in Eq.~(\ref{case 1}), which
is allowed by conservation of angular momentum. The energy
denominators and the function $Q$ remain unchanged because $|
f(k_t,\widetilde{m})|=|f(k_t,m)|$, while the sign before $\Delta \mu
q$ changes. Notice that for A-SWNTs, $f(k_t,\widetilde{m})=e^{i
4\pi/3}f^*(k_t,m)$, so that by the definition of $\phi$,
$\cos(\widetilde{\phi}-\widetilde{\phi'})=\s\s'\cos(\phi-\phi')$.
Define $\widetilde{\Psi}_i\equiv \Psi_{\s_i}(k_t,2n-m_i)$, then
\begin{eqnarray}\label{Ppp*:reflection}
 P_{\pi\pi^*}(\{\widetilde{\Psi}_i\})&=&\left(\prod\limits_{i=1}^{\mu}\s_{i-1}\s_{i}\right)
P_{\pi\pi^*}(\{\Psi_i\})\nonumber\\ &=&\s_{\pi}\s_{\pi^*}
P_{\pi\pi^*}(\{\Psi_i\})=-P_{\pi\pi^*}(\{\Psi_i\}),
\end{eqnarray}
\noindent which leads to
\begin{equation}\label{Hpp*:reflection}
H^{(\mu)}_{\pi\pi^*}(\{\Psi_i\})+H^{(\mu)}_{\pi\pi^*}(\{\widetilde{\Psi}_i\})\propto
\left(\frac{U_q}{2}\right)^{\mu}\sin(\Delta \mu
q\t_0)P_{\pi\pi^*}(\{\Psi_i\}).
\end{equation}
\noindent Since the above relation is true for all possible sets of
intermediate states at all coupling orders, one can conclude that: (a)
The coupling between $\pi$ and $\pi^*$ subbands is \emph{always} zero
if $\t_0=0$, i.e., when a mirror plane of the potential overlaps with
one of the vertical mirror planes (or glide planes) of the
A-SWNT. This is an explicit result of the mirror symmetry
requirement.~\cite{LI2004} (b) The coupling is zero if $\Delta \mu=0$,
which results from reflection symmetry of the energy bands:
$E_{\s}(k_t,m)=E_{\s}(k_t,2n-m)$.  This excludes the possibility of
any nonzero second order contribution, i.e., $\mu=2, \Delta\mu=0$. In
other words, a \emph{second order} band gap is forbidden in
A-SWNTs. (c) The next \emph{lowest possible} $\Delta\mu$ satisfying
the angular momentum conservation in Eq.~(\ref{m_consv}) is
$\mu_0\equiv 2n/\mathrm{gcd}(2n,q)$, which is also the lowest
contributing order of the perturbation series.~\cite{LI2004} For small
$U_q$, the band gap opening will be at least the $\mu_0$-th order in
$U_q$,
\begin{eqnarray}
  E_g\approx2|H_{\pi\pi^*}(\kf)|\sim
  \frac{\vf}{R}u^{\mu_0}\sin(\mu_0q\t_0)h(q,n)+(\mbox{terms of
  $\mu>\mu_0$}),
\end{eqnarray}\noindent where $\kf=2\pi/3a$ is the Fermi point with
$a\simeq 2.5$ \AA. $u=U_qR/\vf$ is the dimensionless potential and
$h(q,n)$ is a complicated function depending on the angular momentum
of the potential, $q$, as well as the A-SWNT index, $n$.

The summation over all possible intermediate states can be further
simplified by combining the original process with $\{\Psi_i\}$ and the
reversal process with
$\{\Psi^R_i\equiv\Psi_{-\s_{\mu-i}}(k_t,2n-m_{\mu-i})\}$. The sign
change of the energy results in an extra factor of $(-1)^{\mu-1}$ in
the denominator and the function $Q$ changes accordingly. It can be
proved that $P_{\pi\pi^*}(\{\Psi^R_i\})=P_{\pi\pi^*}(\{\Psi_i\})$ and
therefore at small $\o$ the relation holds:
\begin{eqnarray}\label{Hpp*:time reversal}
 H^{(\mu)}_{\pi\pi^*}(\{\Psi_i\})+H^{(\mu)}_{\pi\pi^*}(\{\Psi^R_i\})
 &\propto&\left[1-(-1)^{\mu}\right]-\o\left[1+(-1)^{\mu}\right]\sum_{i=1}^{\mu-1}\s_i|f_i|
 +O(\o^2).
\end{eqnarray}
\noindent Within the orthogonal model ($\o=0$), only the first term in
Eq.~(\ref{Hpp*:time reversal}) exists and is nonzero when $\mu_0$, and
therefore $\mu$, is odd.  This constrain on $\mu_0$ results from the
invariance of the inner product of pseudo-spinors upon reversal
operation, in combination with the electron-hole symmetry
$E_{-\s}^0(k_t,m)=-E_{\s}^0(k_t,m)$. The latter, however, is not an
intrinsic property of SWNTs, but rather due to the nearest neighbor
approximation. For instance, the energy band symmetry is broken when
the second nearest neighbor hopping integral is included, or,
equivalently, if $\o\neq0$.  In the latter case, $Q(\{\Psi_i\})$ and
$-Q(\{\Psi^R_i\})$ don't cancel out and a non-zero band gap
proportional to $\o$ opens for even $\mu_0$.  Figure~\ref{fig:Egsin2t}
plots the variation of band gap at $q=2$ calculated within the
orthogonal and nonorthogonal TB models. At $s=0$, the (6,6) A-SWNT
remains metallic, because the coupling order $\mu_0=6$ is
forbidden. At nonzero $\o$, a small band gap occurs and increases as a
power law of $u$. It is also found that the band gap grows linearly
with the magnitude of $\o$, consistent with the prediction of
Eq.~(\ref{Hpp*:time reversal}). In contrast, the band gap curve of a
(5,5) A-SWNT only shows a slight increase at nonzero $\o$, because the
corrections is of the order of $\o^2$.

\subsection{Diagonal coupling: renormalization of the Fermi velocity}

Except for a few special cases, for instance with $q=\mbox{multiples
of } 2n$, the coupling order between $\pi$ and $\pi^*$ subbands is
about the same order of $n$ and the resulting band gap remains
small. For example, $\mu_0=2n$ at $q=1$, and $\mu_0=n$ at $q=2$.
However, the diagonal coupling matrix elements $H_{\pi\pi}$ and
$H_{\pi^*\pi^*}$ are not necessarily small. Same symmetry arguments
can be applied here. Upon reflection or reversal operation,
$P_{\pi\pi}$ (or $P_{\pi^*\pi^*}$) remains the same and one obtains
\begin{eqnarray}
H^{(\nu)}_{\alpha\alpha}(\{\Psi_i\})&+&H^{(\nu)}_{\alpha\alpha}(\{\widetilde{\Psi}_i\})\propto
 \left(\frac{U_q}{2}\right)^{\nu}\cos(\Delta \nu
 q\t_0)P_{\alpha\alpha}(\{\Psi_i\})\label{Hpp:reflection}\\
 H^{(\nu)}_{\alpha\alpha}(\{\Psi_i\})&+&H^{(\nu)}_{\alpha\alpha}(\{\Psi^{\st
 R}_i\})\nonumber\\ & &\propto\left[
 P_{\alpha\alpha}-(-1)^{\nu}P_{\beta\beta}\right]-\o\left[P_{\alpha\alpha}+(-1)^{\nu}P_{\beta\beta}\right]\sum_{i=1}^{\nu-1}\s_i|f_i|+O(\o^2)\label{Hpp:time
 reversal},
\end{eqnarray}
\noindent where $\alpha(\beta)=\pi,\pi^*$ and $\alpha\ne\beta$.  Since
$\cos(\Delta\nu q\t_0)$ is always unity whenever $\Delta \nu=0$ , the
lowest contributing coupling order is therefore $\nu=2$. So unlike the
off-diagonal coupling, nonzero diagonal terms $H_{\pi\pi}$ and
$H_{\pi^*\pi^*}$ can always be obtained from low order angular
perturbation. In an orthogonal basis ($\o=0$), only the first term in
Eq.~(\ref{Hpp:time reversal}) remains, which corresponds to an energy
shift for $\pi$ and $\pi^*$ subbands in the same direction when
$\nu=\mbox{odd}$ and the opposite direction when $\nu=\mbox{even}$. If
$\o\neq0$, relative shift between the two subbands always occurs. 
The values of $H_{\pi\pi}$ and $H_{\pi^*\pi^*}$ do not contribute to
the band gap opening of the A-SWNT, but may influence the Fermi point
position as well as the DOS near the Fermi level. Assume that $\o=0$,
then the second order perturbation summation is reduced to
\begin{eqnarray}
  \sum_{\{\Psi_i\}}
H_{\pi\pi}^{(2)}&=&2\left(\frac{U_q}{2}\right)^2\frac{\cos^2\left[\phi_{\pi}-\phi_+(k_t,n+q)\right]-\cos^2\left[\phi_{\pi^*}-\phi_+(k_t,n+q)\right]}{-E_+(k_t,n+q)}\nonumber\\
&=&\frac{U_q^2}{2|\g_0|}F(k_t,q)=-\sum_{\{\Psi_i\}}H_{\pi^*\pi^*}^{(2)}\nonumber\\
F(k_t,q)&\approx&-\frac{1}{2\left[1+2\cos(q\pi/n)\right]}+\frac{\sqrt{3}\cos(2q\pi
/3n)}{8\sin^2(q\pi/2n)}(|k_t|-\kf)a+O\left(\Delta k^2\right).
\end{eqnarray}
\noindent
The $\pi$ subband is a \emph{decreasing} function of $|k_t|$, and
since $F(k_t,q)$ is an \emph{increasing} function of $|k_t|$ at small
$q$ values, it becomes flattened near $\pm \kf$ as a result of the
second order perturbation.  Similar trend can be found for the $\pi^*$
subband.  The new Fermi points move toward $k_t=0$ as $F(\pm
\kf,q)<0$, and the renormalized Fermi velocity is given by

\begin{equation}
\bar{v}_F\approx\left(1+\frac{u^2}{2q^2}\right)^{-1}
\left(\vf-\frac{\sqrt{3}n^2a}{4q^2\pi^2}\frac{U_q^2}{|\g_0|}\right)\approx\left(1-\frac{u^2}{q^2}\right)\vf,
\end{equation}\noindent
where the prefactor $(1+u^2/2q^2)^{-1}$ is due to the normalization of
the perturbed wave function.  $\bar{v}_F$ can also be estimated from a
chiral gauge transformation~\cite{NOVI2002} as $J_0(2u/q)\vf\approx
(1-u^2/q^2)\vf$, with $J_0$ the Bessel function of the first kind. The
renormalized Fermi velocity of a $(10,10)$ A-SWNT is plotted in
Fig.~\ref{fig:vF}, and excellent agreement is found between the TB
results and the analytical predictions. Under potentials of small
$q$'s, the shape of $\pi$ and $\pi^*$ subbands is strongly perturbed
and the low energy DOS is greatly enhanced (see insets of
Fig.~\ref{fig:vF}), which becomes more evident for large radius
A-SWNTs even at a relatively weak perturbation.

One may note that, for A-SWNTs, the finite curvature shifts the Fermi
point further toward $k=0$ while inclusion of $\sigma$-orbitals may
also modify the magnitude of the band gap
opening.~\cite{HAMA92,BLAS94} For the radius range considered in this
paper, the correction remains small, and since our symmetry arguments
are based on the lattice geometry of A-SWNTs (mirror or angular
symmetry), the selections rules are not affected.

\section{Tensor Potential: $q=2$}\label{Sec:tensor potential}
 Another type of perturbation is realized by changing the hopping
integrals between neighboring atomic orbitals, for example, by elastic
deformations. Due to the high flexibility, carbon nanotubes can
sustain remarkable deformations, which cause drastic changes in the
electronic properties.  Stretching, twisting and squashing the
nanotubes have been attempted both theoretically \cite{KANE97,LIU99,
PARK99,GULS2002a,LU2003,MEHR2005,ROCH99,CHAR2001} and
experimentally.~\cite{CAO2003,MINO2003,LILJ2004, WU2004} Here, we
apply a radial deformation across the A-SWNT and derive the condition
to open a secondary band gap.

Assume that a uniform stress is applied along $y$ direction, which is
rotated from the A-SWNT's vertical mirror planes (or glide planes) by
$\t_0$. The cross section of the A-SWNT is distorted into an ellipse
with the two axes given by
\begin{equation}\label{deformation}
  R_y=R_0(1-\e),\hspace{.2 in}  R_x=R_0(1+\nu\e),
\end{equation}
\noindent where $\e$ is the strain along $y$ direction and $\nu$ is
the cross section Poisson ratio. At small strain, the hopping integral
between nearest neighbors can be linearly expanded as
\begin{eqnarray}\label{Dgamma}
\g_{\lambda,i}-\g_0&\approx&\left(\frac{\partial\g}{\partial
  r}\right)_{r_0}\Delta
  r_{\lambda,i}=\delta\g_{\lambda}\left(\frac{1-\nu}{1+\nu}+\cos2\t_{\lambda,i}\right)\\
  \delta\g_{\lambda}&=&-2\e r_0(1+\nu)
  \left(\frac{R_0}{r_0}\sin\frac{\t_{\lambda}}{2}\right)^2\left(\frac{\partial
  \g}{\partial r}\right)_{r_0},
\end{eqnarray}
\noindent where $\t_{\lambda,i}$ corresponds to the midpoint of the
$i$-th bond along ${\bf r}_{\lambda}$ direction. The first term in
bracket in Eq.~(\ref{Dgamma}) has no angular dependence and does not
break mirror symmetry
because $\t_2=\t_3$ and thus $\delta\g_2=\delta\g_3$. Since the
Poisson ratio was found to be close to unity,~\cite{GULS2002a} the
second term
dominates, which corresponds to an angular momentum $q=2$.

The perturbation matrix element $H_{\alpha\beta}$ can be expanded into
a perturbation series in a similar fashion as in Eq.~(\ref{case 1}),
but now with $Q_{\alpha\beta}(\{\Psi_i\})=1$ and
\begin{eqnarray}
    P_{\alpha\beta}(\{\Psi_i\})&=&\prod\limits_{i=1}^{\mu}\sum_{\lambda=1}^3g_{\lambda}(k_t;m_{i-1},\s_{i-1};m_i,\s_i)
\end{eqnarray}
\noindent where $g_{\lambda}$ is defined in
Eq.~(\ref{Eq.vector}). Using the fact that $\delta\g_2=\delta\g_3$,
one can prove that
\begin{eqnarray}\label{g sum}
\sum_{\lambda=1}^3g_{\lambda}(k_t;m,\s_;m',\s')
&=&\s\s'\sum_{\lambda=1}^3g_{\lambda}(k_t;2n-m,\s_;2n-m',\s'),\label{tensor_reflection}\\
&=&-\sum_{\lambda=1}^3g_{\lambda}(k_t;2n-m',-\s';2n-m,-\s)\label{tensor_reversal}.
\end{eqnarray}
\noindent By applying the reflection operation $\Psi \to
\widetilde{\Psi}$ on the intermediate states,
Eq.~(\ref{Ppp*:reflection}) and (\ref{Hpp*:reflection}) are recovered,
as required by mirror symmetry conservation. The lowest contributing
order is therefore $\mu_0=2n/\mathrm{gcd}(2n,2)=n$. Now apply a
reversal operation $\Psi \to \Psi^R$ as in Sec.~\ref{Sec:scalar
potential}. Since $E^0_{-\s}(k_t,2n-m)=-E^0_{\s}(k_t,m)$, from
Eq.~(\ref{tensor_reversal}) one has
\begin{equation}
   H^{(\mu)}_{\pi\pi^*}(\{\Psi_i\})+H^{(\mu)}_{\pi\pi^*}(\{\Psi^{\st
   R}_i\})= H^{(\mu)}_{\pi\pi^*}(\{\Psi_i\})\left[ 1+
   \frac{(-1)^{\mu}}{(-1)^{\mu-1}}\right]=0.
\end{equation}
This means that a band gap does not occur in any perturbation order
and such strain cannot induce MST in A-SWNTs.
Certainly, a hidden symmetry, namely, the electron-hole symmetry
$E^0_{-\s}(k_t,-m)=-E^0_{\s}(k_t,m)$, forbids the $\pi$ and $\pi^*$
subband mixing. By including the second nearest neighbor interactions,
this symmetry can be weakly broken and a finite band gap
occurs.~\cite{g1} The magnitude of this band gap depends strongly on
the parity of the A-SWNT index $n$. It was reported earlier that
squashing a $(6,6)$ or $(8,8)$ A-SWNT does not induce a MST in the
range of elastic deformation,~\cite{GULS2002a,LU2003} and the nanotube
remains metallic until the two opposite walls are brought close enough
to form new bonds. We prove that the vanishing band gap is caused by
the high coupling order between $\pi$ and $\pi^*$ subbands, $\mu_0=n$,
and the additional smallness of the overlap integral $\o$ and higher
neighbor interactions, which make it impossible to observe the MST
effect until the A-SWNT collapses. The situation is quite different,
for example, for a $(5,5)$ A-SWNT which has an odd number index and a
smaller coupling order, and a finite band gap was observed at moderate
deformation.~\cite{PARK99} On the other hand, when the radial
deformation is large enough to induce strong $\pi$-$\sigma$
interaction, the single $\pi$-orbital description may no longer be
sufficient.

\section{Combination of different types of potentials}\label{Sec:combination}
One way to reduce the coupling order $\mu_0$ is by combining
potentials of different angular momentum. For example, we have shown
that by applying a scalar potential of the form of
$V_0\left(\sin\t+\sin 2\t\right)$, $\mu_0$ can be reduced to three for
all values of $n$.~\cite{LI2004} The combination of elastic radial
deformation and uniaxial electrostatic potential will have a similar
effect. By choosing appropriate angular momentum and relative position
of the two components, the coupling order can be even lowered to
$\mu_0=2$, as shown below.

Assume that a scalar potential (denoted as $U$) is applied on an
A-SWNT together with a tensor perturbation (denoted as ${\cal
E}\propto \delta\g_q$) of same angular momentum $q$, but with an
angular difference $\t_d$ between the mirror planes of these two
components. As shown in Sec.~\ref{Sec:scalar potential} and
Sec.~\ref{Sec:tensor potential}, the second order contribution from
either component is zero, but the cross terms do not necessarily
vanish. For the second order coupling between $\pi$ and $\pi^*$
subbands, $\Psi_\pi \stackrel{{\cal
E},U}{\longrightarrow}\Psi_{\s}(k_t,m)\stackrel{U,{\cal
E}}{\longrightarrow}\Psi_{\pi^*}$, there are 8 different cross terms
with $m=n\pm q$ and $\s=\pm 1$. At $k_t=\kf$, the cross terms adds up
to
\begin{eqnarray}\label{combination}
\sum H^{(2)}_{\pi\pi^*,\st{cross}}(\kf)&= &\sin (q\t_d)
\sum_{\lambda=1}^3\frac{U_q\delta\g_{\lambda,q}\cos(\kf z_{\lambda})}{i|E^0_+(\kf,n+q)|}\sin\left[2\phi_+(\kf,n+q)-(n+q/2)\t_{\lambda}\right]\nonumber\\
&\propto&u\e\sin (q\t_d)\frac{|\g_0|}{q}, \qquad q\ll n,
\end{eqnarray}
\noindent where $u=U_qR/\vf$ and $\e\sim\delta\g/\g_0$ are the
dimensionless potential and strain respectively. According to
Eq.~(\ref{combination}), the coupling between $\pi$ and $\pi^*$
subbands is largest when $\t_d=\pi/2q$ and always vanishes whenever
the mirror planes of $U$ and ${\cal E}$ overlap. The magnitude of the
band gap, $E_g\approx2|\sum H^{(2)}_{\pi\pi^*,\st{cross}}(\kf)|$, is
linear in both $u$ and $\epsilon$, and the dependence on the A-SWNT
radius $R$ (or index $n$) is very weak. This differs from the
situation of mixed scalar potentials of different angular momentum, in
which case $E_g$ decreases with $R$ as an inverse power
law.~\cite{LI2004}

In the case of radial deformation (Sec.~\ref{Sec:tensor potential}),
the tensor perturbation on the hopping integrals and the effective
on-site potential~\cite{g1} have overlapping mirror planes
so that no second order coupling occurs.  Assume that one applies on
the A-SWNT a scalar potential with the $q=2$ component shifted
$\t_d=\pi/4$ relative to the stress, for example by changing the
electrostatic environment around the A-SWNT. The resulting band gap is
plotted in Fig.~\ref{fig:Eg_comb} for A-SWNTs of different radius as a
function of $u$. The hopping integral under deformation is assumed to
change as $\g\propto r^{-2}$, where the new bond length $r$ is
calculated from Eq.~(\ref{deformation}). The Poisson ratio is taken to
be unity. The numerical values of $E_g$ clearly follow a linear
dependence on $u$ and $\e$ when the perturbation is weak.  The radius
dependence is barely seen even at large $u$, which means that one can
always generate a substantial band gap in a large radius A-SWNT using
only a moderate external potential.  Recently, it was found that the
conductance of a carbon nanotube can be controlled by tuning the
voltage of the local gate placed near mechanical defects on the tube
(kinks or bends).~\cite{BIER2004} This scenario is reminiscent of the
combination of scalar and tensor potentials of $q=1$, and might be
related to the resulting second-order band gap.

\section{Metallic SWNTs with arbitrary chirality}\label{Sec:chiral}

In this section, we generalize our derivation of subband coupling to
arbitrary metallic SWNTs by expanding the TB wave functions near the
Fermi point. Only angular perturbations will be considered here, i.e.,
the axial wave vector is always conserved.  An analogy is made to the
cancellation rules of the back scattering process in
SWNTs.~\cite{ANDO} For the sake of clarity of the derivation, the
overlap integral $\o$ is assumed zero. We remind that some
cancellation rules will be weakly broken if the electron-hole symmetry
is lifted, for example, by $\o\neq 0$.

First, the electronic states are approximated as the product of a
plane wave part
and a pseudo-spinor part by expanding the wave vector near the Fermi
Point $\bf{K}$ of two-dimensional graphite:
 \begin{eqnarray}\label{def:phi2}
\Psi_{\s}(\hat{\bf{k}})&=&\frac{e^{i\hat{\bf{k}}\cdot\bf{r}}}{\sqrt{2}}\left(\
\begin{array}{r}
e^{i\phi_{\hat{\bf{k}},\s}}\\
e^{-i\phi_{\hat{\bf{k}},\s}}
\end{array}\right),
\hspace{.2 in}2\phi_{\s}(\hat{\bf
k})=-\s\frac{\pi}{2}+\arg{(\hat{k}_t+i\hat{k}_c})+\eta,
\end{eqnarray}
\noindent where $\hat{\bf{k}}=(\hat{k}_t,\hat{k}_c=\hat{m}/R)$ is
measured from ${\bf K}$ and $\eta$ is the chiral angle. It can be
proved that the definition of $\phi_{\s}(\hat{\bf k})$ here is
consistent with that of $\phi_{\s}({\bf k})$ in Sec.~\ref{Sec:matrix
element}.  In contrast to Ref.~\onlinecite{ANDO}, the phase difference
due to the sign of $\s$ is absorbed in the definition of
$\phi_{\s}(\hat{\bf k})$ so that the product of two pseudo-spinors is
always real. In addition, the wave functions of the two crossing
subbands of metallic SWNTs remain continuous as a function of
$\hat{k}_t$ when the wave vector passes through the Fermi point.

Assume applying an arbitrary angular scalar potential
$V(\t)=\sum_{q}V_qe^{iq\t}$, where $V_q$ is the angular Fourier
component of the potential: $V_q \equiv(2\pi)^{-1} \oint d\t
V(\t)e^{-iq\t}$.  The \emph{direct} coupling matrix element between
two states $\Psi_{\s_1}(\hat{\bf k})$ and $\Psi_{\s_2}(\hat{\bf k}')$,
with $\hat{k}_{t}=\hat{k}'_{t}$, can be reduced to
\begin{eqnarray}\label{scalar potential}
M(\hat{\bf k},\s;\hat{\bf k}', \s')&\equiv&\bra{\hat{\bf
k},\s}V(\t)\ket{\hat{\bf k}',\s'}\nonumber\\
&\approx&\sum_QU_{\hat{m}-\hat{m}'+Q}\nonumber\\
&\times&e^{iQ(\t_{0A}+\t_{0B})/2}\cos\left[\phi-\phi'+Q(\t_{0B}-\t_{0A})/2\right],
\end{eqnarray}
\noindent where $\t_{0A}$ and $\t_{0B}$ are angular coordinates of any
$A$ and $B$ atoms. Angular quantum number $Q$ corresponds to the
angular analog of the reciprocal lattice vector of two-dimensional
graphite and accounts for contributions from the short wave-length
component of the potential. When all nonzero $Q$'s are neglected, the
matrix element $M(\hat{\bf k},\s;\hat{\bf k}', \s')$ is reduced to the
product of the Fourier transform of the potential and the inner
product of two pseudo-spinors, comparable with the on-site term in
Eq.~(\ref{Eq.matrix element}) for A-SWNTs.

If the perturbation is a tensor potential, i.e., affecting the
off-site hopping integrals instead of the on-site energies, one can
similarly write the coupling matrix as:
\begin{eqnarray}\label{tensor potential}
M(\hat{\bf k},\s;\hat{\bf k}', \s')&\equiv&\bra{\hat{\bf
k},\s}\delta\gamma^{\st{op}}\ket{\hat{\bf k}',\s'}\nonumber\\
&\approx&\sum_{\lambda=1}^3\delta
\gamma_{\lambda,\hat{m}-\hat{m}'}\nonumber\\
&\times&\cos\left[\phi+\phi'-\left(\frac{\hat{\bf k}+\hat{\bf
k}'}{2}+{\bf K}\right)\cdot{\bf r}_{\lambda}\right],
\end{eqnarray}
\noindent where $\delta \gamma_{\lambda,q}$ is the discrete angular
Fourier transform of the change in hopping integrals: $\delta
\gamma_{\lambda,q}=N^{-1}\sum_i\delta\gamma_{\lambda,i
}e^{-iq\t_{\lambda,i}} $.  Comparing with Eq.~(\ref{scalar
potential}), one finds that matrix elements for the two types of
perturbations have different dependence on the phase angle
$\phi_{\s}(\hat{\bf k})$.

\subsection{First-Order Coupling}
The first-order subband coupling corresponds to direct mixing between
$\Psi_{+}(\hat{\bf k})$ and $\Psi_{-}(\hat{\bf k})$ and has
straightforward description within nearly degenerate perturbation
theory, where $\hat{\bf k}=(\hat{k}_t,0)$ for metallic SWNTs.  Since
$\phi_{\s}(\hat{\bf k})$ is continuous in vicinity of $\hat{k}_t=0$,
the diagonal matrix element $M(\hat{k}_t,\s;\hat{k}_t,\s)$ is
continuous as well and merely shifts the location of the Fermi point
and renormalizes the Fermi velocity.  The change of the band gap is
therefore determined by the off-diagonal term
$M(\hat{k}_t,\s;\hat{k}_t,-\s)$.

According to Eq.~(\ref{scalar potential}), only Fourier components of
the scalar potential with $q=Q$'s contribute to the direct
coupling. For example, for $(n, n)$ armchair or $(n, 0)$ zigzag
nanotubes, it can be an angular perturbation of the form $
\cos{(2n\t)}$. Such Fourier component can be obtained by applying
torsion, using chemical/biological decoration of the tube surface or
the high multipoles of inhomogeneous potentials.  On the other hand,
since the pseudo-spinors of $\Psi_{+}({\bf k})$ and $\Psi_{-}({\bf
k})$ are always orthogonal, the matrix element in Eq.~(\ref{scalar
potential}) is reduced to
\begin{eqnarray}\label{reduced scalar potential}
M(\hat{k}_t,+;\hat{k}_t,-)&=&\sum_QU_Qe^{iQ(\t_{0A}+\t_{0B})/2}\sin\frac{Q(\t_{0B}-\t_{0A})}{2}\nonumber\\
&=&i\left(\widetilde{U}_A-\widetilde{U}_B\right)/2,
\end{eqnarray}
\noindent with $\widetilde{U}_{A,B}=\sum_QU_Qe^{iQ\t_{0A,0B}}$.
Eq.~(\ref{reduced scalar potential}) indicates that such potential
components must be distinguishable at the two sublattices so as to mix
the two orthogonal pseudo-spinors directly.

For the tensor potential, the symmetry is lower and even a uniform
deformation ($q=0$) can result in the first order coupling. The matrix
element in Eq.~(\ref{tensor potential}) is reduced to:
\begin{eqnarray}
M(\hat{k}_t,+;\hat{k}_t,-)&=&\sum_{\lambda=1}^3\delta
\gamma_{\lambda}\cos\left[\arg(\hat{k}_t)+\eta-(\hat{{\bf k}}+{\bf
K})\cdot{\bf r}_{\lambda}\right]\nonumber\\
&\stackrel{\hat{k}_t\rightarrow
0}{=}&\mathrm{sign}(\hat{k}_t)\left[\delta\g_1\cos(\eta-2\pi/3)+\delta\g_2\cos\eta+\delta\g_3\cos(\eta+2\pi/3)\right]\nonumber\\
&=&\mathrm{sign}(\hat{k}_t)\left[\cos\eta\left(\delta\g_2-\frac{\delta\g_1+\delta\g_3}{2}\right)+\frac{\sqrt{3}}{2}\sin\eta\left(\delta\g_1-
\delta\g_3\right)\right],
\end{eqnarray}
\noindent and more specifically for metallic achiral SWNTs,
\begin{eqnarray}
M(0,+; 0,-)&\sim&\left\{ \begin{array}{lr} \delta \gamma_2-\delta
  \gamma_3, & \mbox{ armchair}\\ 2\delta \gamma_2-(\delta
  \gamma_1+\delta \gamma_3), &\mbox{ zigzag}
  \end{array}\right.,
\end{eqnarray}
\noindent which is consistent with previous findings about band gap
changes in metallic SWNTs under uniaxial and torsional
strain.~\cite{LIU99}

\subsection{High-Order Coupling}

When the first order coupling between states $\Psi_{+}(\hat{\bf k})$
and $\Psi_{-}(\hat{\bf k})$ is forbidden, one has to turn to higher
orders of the perturbation.  Deriving the coupling could be tedious
but some general rules can be built using the symmetry of
pseudo-spinors. Similar to the case of A-SWNTs, coupling between the
two crossing subbands of arbitrary metallic SWNT can be represented by
$H_{+-}(\hat{k}_t)=\sum_{\mu}\sum_{\{\Psi_i\}}H_{+-}^{(\mu)}(\{\Psi_i\})$,
and the lowest contributing coupling order $\mu_0$ can be determined
by the dominating Fourier components of the potential.

We first discuss the case of a scalar potential $V(\t)$ and restrict
$Q=0$ in Eq.~(\ref{scalar potential}). The cancellation rule is
similar to those which led to Eq.(\ref{Hpp*:time reversal}). Assume
there is a $\mu$-th order coupling process between $\Psi_{+}(\hat{\bf
k})$ and $\Psi_{-}(\hat{\bf k})$ through intermediate states
$\{\Psi_i\equiv\Psi_{\s_i}(\hat{k}_t,\hat{m}_i)\}$ with
$i=1\cdots\mu-1$:
\begin{eqnarray}\label{original process}
H_{+-}(\{\Psi_i\})&=&\frac{\prod\limits_{i=1}^{\mu}U_{\hat{m}_{i-1}-\hat{m}_i}\cos\left[\phi_{\s_{i-1}}(\hat{k}_t,\hat{m}_{i-1})-\phi_{\s_i}(\hat{k}_t,\hat{m}_i)\right]}{\prod\limits_{i=1}^{\mu-1}\left[-E^0_{\s_i}(\hat{k}_t,\hat{m}_i)\right]},
\end{eqnarray}
\noindent where subscripts ``$0$'' and ``$\mu$'' correspond to
$\Psi_{+}(\hat{\bf k})$ and $\Psi_-(\hat{\bf k})$ respectively, with
$\hat{m}_0=\hat{m}_{\mu}=0$.  define a reversal process with
intermediate states
$\{\Psi^R_i\equiv\Psi_{-\s_{\mu-i}}(\hat{k}_t,-\hat{m})\}$.  After
rearrangement of the summation orders and using the relation
$\phi_{-\s}(\hat{k}_t,-\hat{m})=\eta-\phi_{\s}(\hat{k}_t,\hat{m})$,
one arrives at
\begin{eqnarray}\label{Hab:reversal}
H_{+-}(\{\hat{k}_t,-\hat{m}_{\mu-i},-\s_{\mu-i}\})&=&\frac{\prod\limits_{i=1}^{\mu}U_{\hat{m}_{i-1}-\hat{m}_i}\cos\left[\phi_{-\s_{i-1}}(k_t,-m_{i-1})-\phi_{-\s_i}(\hat{k}_t,-\hat{m}_i)\right]}{\prod\limits_{i=1}^{\mu-1}\left[-E^0_{-\s_i}(\hat{k}_t,-\hat{m}_i)\right]}\nonumber\\
&\approx&(-1)^{\mu-1}H_{+-}(\{\hat{k}_t,\hat{m}_i,\s_i\}),
\end{eqnarray}
\noindent which cancel out with Eq.~(\ref{original process}) for even
$\mu$. The approximation sign for subband coupling process in
Eq.~(\ref{Hab:reversal}) arises from assumptions on the reflection
symmetry $E^0_{\s}(\hat{k}_t,\hat{m})=E^0_{\s}(\hat{k}_t,-\hat{m})$
and reversal symmetry
$E^0_{\s}(\hat{k}_t,\hat{m})=-E^0_{-\s}(\hat{k}_t,\hat{m})$, which can
be derived from the linear dispersion approximation, i.e.,
$E^0_{\s}(\hat{\bf k})=\s \vf|\hat{\bf k}|$. More generally, the
reflection symmetry does not hold except for A-SWNTs due to the
trigonal warping effect. For example, $E_{\s}(\hat{k}_t,1)\neq
E_{\s}(\hat{k}_t,-1)$ for metallic zigzag SWNTs, and a secondary band
gap $E_g\propto R^{-2}$ always opens under a uniform electric field
perpendicular the nanotube radius.~\cite{LI2003} Here $\hat{m}=\pm1$
is measured relative to the Fermi point ${\bf K}$.  When a nonzero
overlap $s$ or high order nearest neighbor interaction is included,
the reversal symmetry can also be weakly broken. We conclude that the
selection rules for arbitrary metallic SWNTs are similar to those for
A-SWNTs, but may acquire a chirality dependence beyond the linear
dispersion approximation.

For general tensor potentials, $M(\hat{k}_t;\hat{m}_i,\s_i;\hat{m}_j,
\s_j)$ and $M(\hat{k}_t,-\hat{m}_j,-\s_j;-\hat{m}_i, -\s_i)$ usually
have different magnitude and no simple cancellation rule can be
built. An exception is when $\gamma_2=\gamma_3$ for A-SWNT, e.g.,
under a radial deformation. In that case, the coupling between $\pi$
and $\pi^*$ subbands is reduced to zero as shown in
Sec.~\ref{Sec:tensor potential}, due to cancellation from the reversal
process within the nearest neighbor approximation.

\section{Conclusion}\label{Sec:conclusion}

In this paper we employ the group theory approach to clarify the issue
of MST in armchair and other metallic SWNTs under angular
perturbations. We study symmetry requirements of MST on the nanotube
and the perturbation potential, and demonstrate that the smallness of
the MST effect is related to the symmetry of the pseudo-spinor
components of the electron wave functions.  Namely, the spinors of the
crossing subbands are orthogonal, thus, any interaction between them
is strongly weakened. For A-SWNTs, the gap is diminishing for almost
any pure angular perturbation with a \emph{single} angular Fourier
component, due to the high coupling order. The coupling order is
proportional to the number of atoms along the tube circumference for
both types of perturbation studied here: the on-site (scalar)
potential and off-site (tensor) deformation. The MST effect can be
greatly enhanced by combining perturbations of different types and/or
different angular momentums.

We formulate selection rules for the band gap opening and its
dependence on the perturbation strength. The combination of the
diagrammatic derivation of interaction matrix elements and group
theory technique allows one to predict the scaling of the band gap on
the potential: $E_g\propto|{\bf V}|^{\mu_0}$, where the scaling
exponent $\mu_0$ can be easily calculated for an arbitrary metallic
SWNT for given symmetry of the potential. Corrections may arise due to
refinement of the model TB Hamiltonian, e.g., the electron-hole
asymmetry, the inclusion of $\sigma$-orbitals and others not
considered here. As an example of such refined model, we calculate the
gap dependence on the overlap integral $\o$ added to the classic
orthogonal TB model.

We present the analytical expression for the renormalization of the
Fermi velocity, which occurs 
even if no MST is observed. The decrease of the Fermi velocity due to
the perturbation is also seen as the enhancement of the DOS close to
the Fermi level.

The MST effect by the SWNT symmetry breaking could have potential
applications for nanoscale electronic and optoelectronic devices.
Additionally, we emphasize the possibility of engineering the nanotube
DOS even when MST is forbidden under given perturbations, which can be
potentially employed for SWNT opticals as well as switching devices.

\begin{acknowledgments}
This work was supported by the ARMY DURINT contract SIT 527826-08 and
NSF grant CCR 01-21616. S.V.R. acknowledges support by start-up fund
and Feigl Scholarship of Lehigh University, and the NSF grant
ECS 04-03489.
\end{acknowledgments}

\appendix*
\section{Nearly degenerate perturbation theory}\label{appendix:NDPT}
When calculating the coupling between two nearly degenerate states
$\Psi_{\alpha}$ and $\Psi_{\beta}$, it is more convenient to treat
them as degenerate states. Since $E_{\alpha}=E_{\beta}=0$ only at the
crossing point, one can include the energy dispersion by redefining
the unperturbed Hamiltonian and the external perturbation, for
$|E_{\alpha}-E_{\beta}|$ being small:
\begin{eqnarray}
  \tilde{H}_0&=&H_0-E_{\alpha}\ket{\Psi_{\alpha}}\bra{\Psi_{\alpha}}-E_{\beta}\ket{\Psi_{\beta}}\bra{\Psi_{\beta}}\nonumber\\
\tilde{H}_1&=&H_1+H_0-\tilde{H}_0.
\end{eqnarray}
\noindent $\Psi_{\alpha}$ and $\Psi_{\beta}$ now become degenerate
states of $\tilde{H}_0$ with $\tilde{E}_{\alpha,\beta}=0$.  Their
original energy difference is absorbed in the perturbation while other
states $\Psi'$ are not affected:
\begin{eqnarray}
\begin{array}{ll}
\bra{\Psi_{\alpha,\beta}}\tilde{H}_1\ket{\Psi_{\alpha,\beta}}=E_{\alpha,\beta},&
  \bra{\Psi_{\alpha}}\tilde{H}_1\ket{\Psi_{\beta}}=\bra{\Psi_{\alpha}}H_1\ket{\Psi_{\beta}}\\
  \bra{\Psi'}\tilde{H}_1\ket{\Psi_{\alpha,\beta}}=\bra{\Psi'}H_1\ket{\Psi_{\alpha,\beta}},&
  \bra{\Psi'}\tilde{H}_1\ket{\Psi''}=\bra{\Psi'}H_1\ket{\Psi''}.
\end{array}
\end{eqnarray}
\noindent As long as $\tilde{H}_1-H_1$ is small, the rearrangement of
$H_0$ and $H_1$ will not affect the results of the perturbation
theory.


\begin{thebibliography}{27}
\bibitem[{g1()}]{g1}
\bibinfo{note}{The deformation also results in an effective on-site potential,
  $V\propto\cos{(2\t)}$, which contributes to the band gap opening as given by
  Eq.~(\ref{Hpp*:reflection}) and Eq.~(\ref{Hpp*:time reversal}) with $\mu_0=n$
  .}
\end{thebibliography}

\newpage
\section*{Figure Captions}
\begin{figure*}[!t]
\centerline{
\includegraphics[width=5 in]{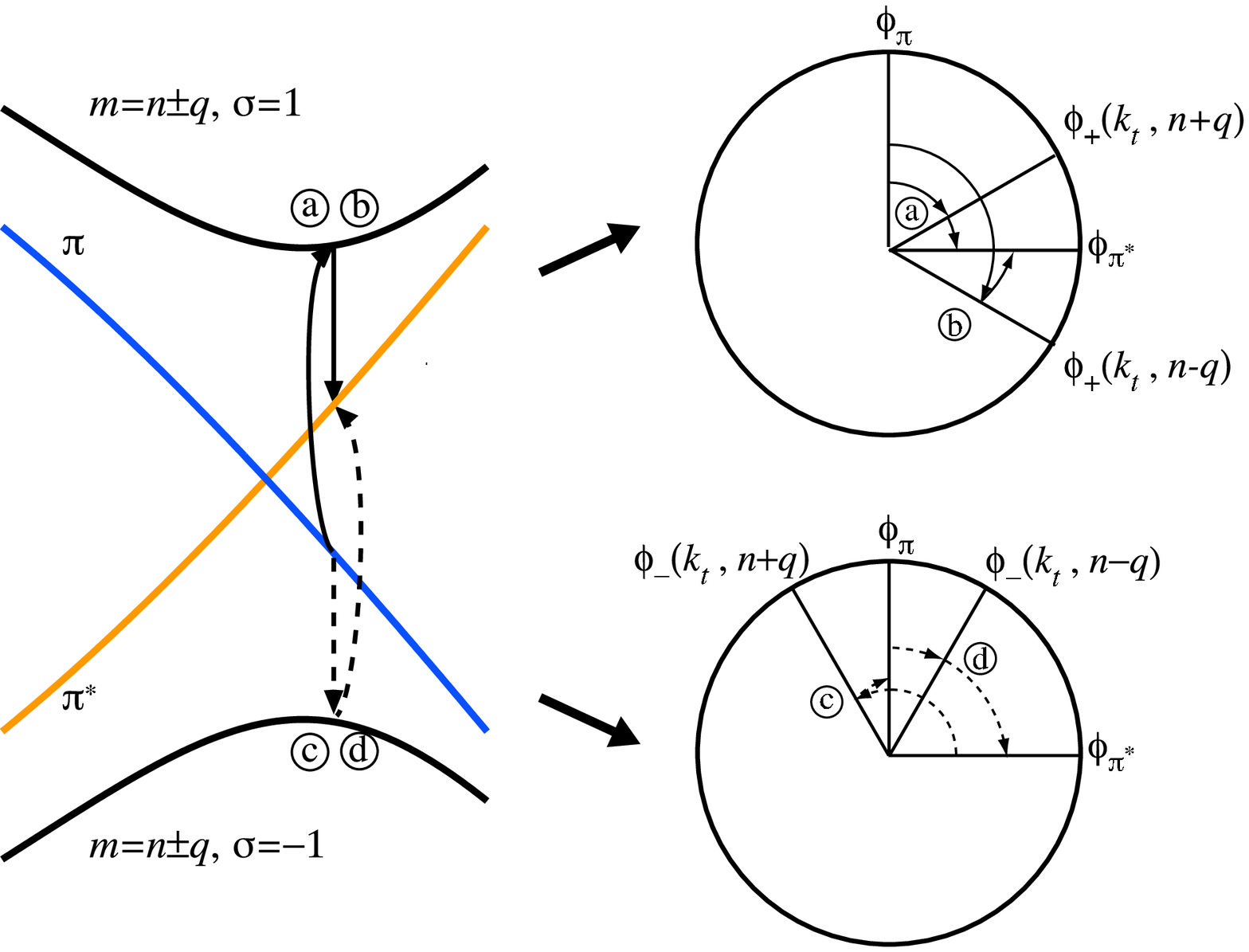}}
\caption{\label{fig:coupling}}
\end{figure*}

\noindent \textbf{Fig.~\ref{fig:coupling}:} (color online). Schematics
  of second order coupling  between $\pi$ and $\pi^*$
  subbands and the corresponding phase angles of intermediate states.
  \\

\begin{figure*}[!b]
\centerline{
\includegraphics[width=6 in]{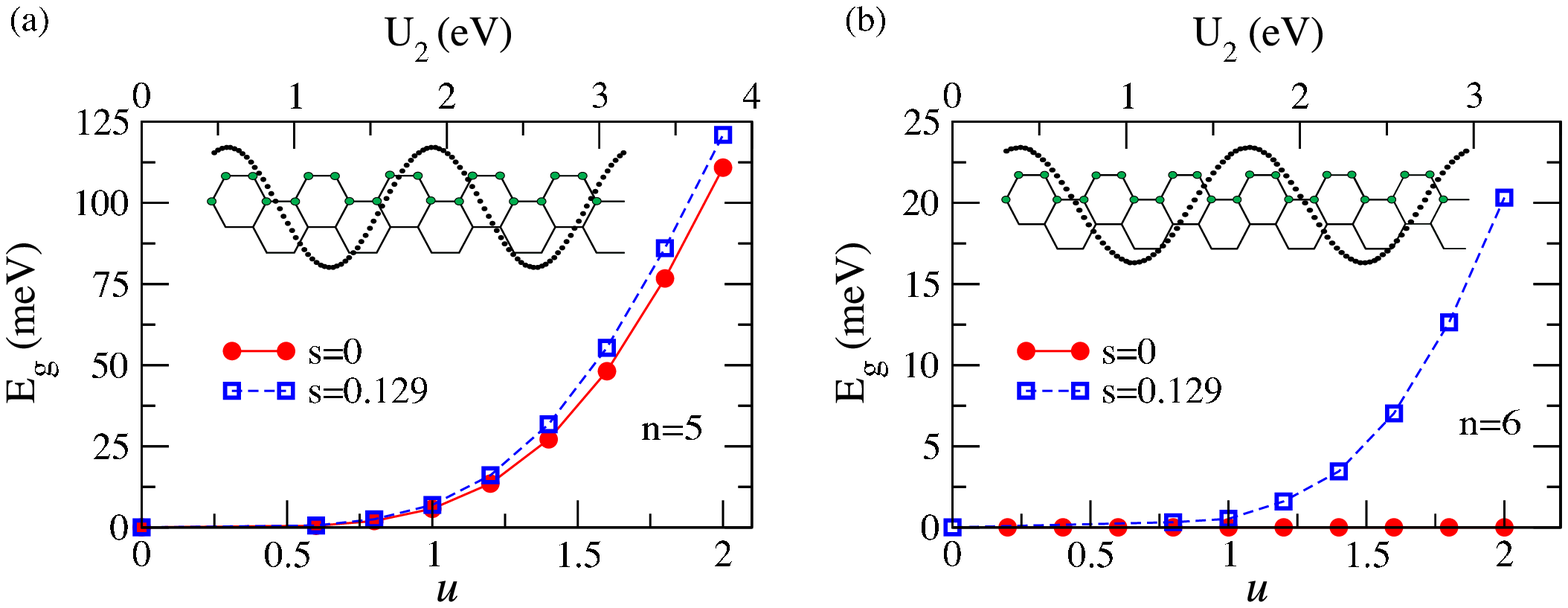}}
\caption{\label{fig:Egsin2t}}
\end{figure*}

\noindent \textbf{Fig.~\ref{fig:Egsin2t}:} (color online).  Band gap
variation of (a) $(5,5)$ and (b) $(6,6)$ A-SWNTs as a function of the
applied angular potential with $q=2$. Insets: the unwrapped unit cell
and schematics of the potential.  Mirror planes of the potential pass
through atomic sites so that all vertical mirror reflection and glide
reflection symmetries are simultaneously broken.  \\

\begin{figure*}[!thb]
\centerline{
\includegraphics[width=4.5 in]{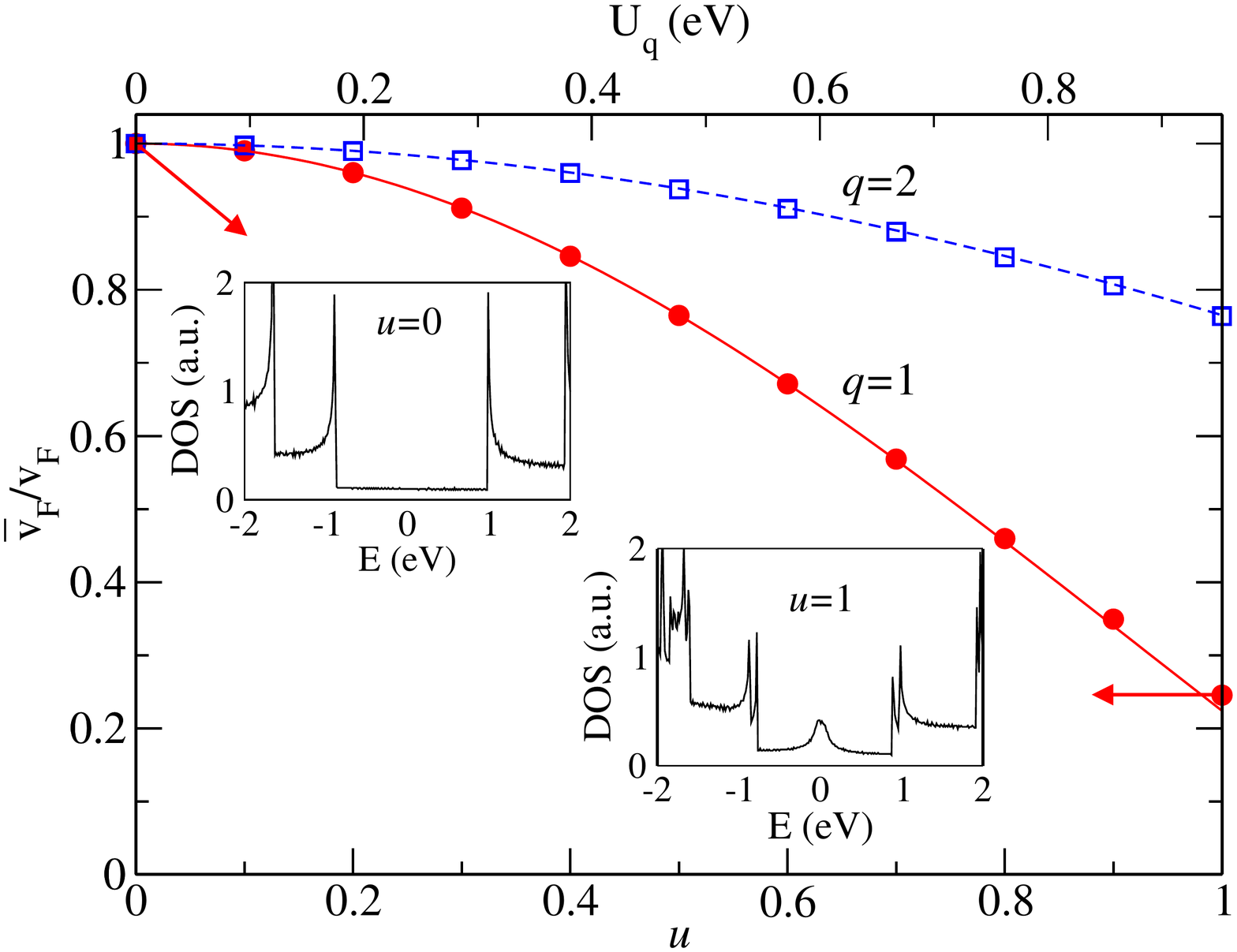}}
\caption{\label{fig:vF}}
\end{figure*}

\noindent \textbf{Fig.~\ref{fig:vF}:} (color online). Renormalized
Fermi velocity $\bar{v}_F$ of a $(10,10)$ A-SWNT as a function of $u$
with $q=1$ (circles) and $q=2$ (squares). Solid and dashed lines are
corresponding predictions from $J_0(2u/q)$. Insets show the DOS
structure near the Fermi level at $u=0$ and $u=1$, with $q=1$.  \\

\begin{figure*}[!b]
\centerline{
\includegraphics[width=4.5 in]{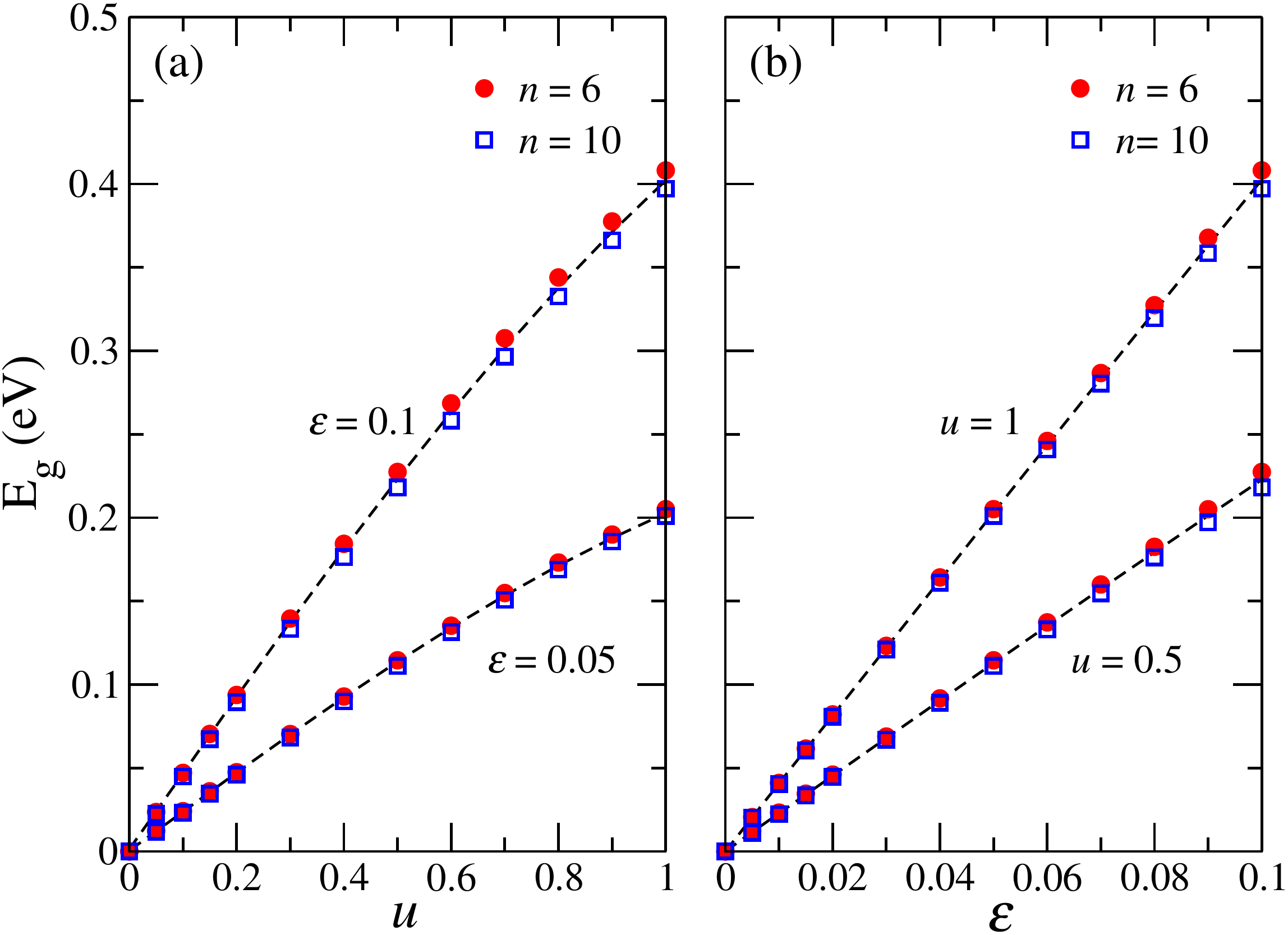}}
\caption{\label{fig:Eg_comb}}
\end{figure*}

\noindent \textbf{Fig.~\ref{fig:Eg_comb}:} (color online). Band gap
  variations of A-SWNTs (a) as a function of $u$ at $\e=$0.05, 0.1,
  and (b) as a function of $\e$ at $u=$0.5, 1. Dashed lines are to
  guide the eye.
\newpage

\end{document}